\documentclass[letterpaper,twocolumn,english,aps, prl]{revtex4}
\usepackage[T1]{fontenc}
\usepackage[latin1]{inputenc}
\usepackage{graphicx}
\usepackage{amssymb}

\makeatletter

\usepackage{graphicx}

\usepackage{amssymb}

\usepackage{floatflt}

\usepackage{times}

\usepackage{babel}

\usepackage{babel}
\makeatother
\begin{document}

\title{Magnetic Flux Periodic Response of Nano-perforated Ultrathin Superconducting
Films}

\date{\today{}}

\author{M. D. Stewart, Jr., Zhenyi Long, James M. Valles, Jr.}

\affiliation{Department of Physics, Brown University, Providence, RI 02906}

\author{Aijun Yin and J. M. Xu}

\affiliation{Division of Engineering, Brown University, Providence, RI 02906}

\begin{abstract}
We have patterned a hexagonal array of nano-scale holes into a series of ultrathin, superconducting Bi/Sb films with transition temperatures 2.65 K $<T_{co} < $5 K. These regular perforations give the films a phase-sensitive periodic response to an applied magnetic field.  By measuring this response in their resistive transitions, $R(T)$, we are able to distinguish regimes in which fluctuations of the amplitude, both the amplitude and phase, and the phase of the superconducting order parameter dominate the transport. The portion of $R(T)$ dominated by amplitude fluctuations is larger in lower $T_{co}$ films and thus, grows with proximity to the superconductor to insulator transition.
\end{abstract}
\maketitle
The superconductors originally considered by BCS exhibited spectacularly
sharp transitions from a finite resistance to zero resistance as a
function of temperature \cite{BCS}. Fluctuation effects were negligible.
Presently, a great deal of attention focuses on low superfluid density
superconductors for which fluctuations strongly influence and substantially
broaden their phase transitions. These include the high temperature
superconductors \cite{Emery&Kivelson-phasechangesTc}, and in particular,
their underdoped versions \cite{Ong-highTcvorticesabove,Maple-highTcvorticesabove,highTc-novorticesabove},
and ultrathin superconducting films near the superconductor to insulator
transition (SIT) \cite{InOx-scale,a-MOGe-Yazdani&Kapitulnik,Goldman-vorticesoninsulatorside,x-overfromphasetoamp-dynes,Shahar-InOx}.
For the latter, resistive transitions, $R(T)$, can develop widths
comparable to or greater than the apparent mean field transition temperature,
$T_{c0}$ \cite{criticalampfluct}.

To discuss the effects of fluctuations on the $R(T)$ it is helpful
to consider the two component superconductor order parameter $\psi=|\psi_{0}|e^{i\phi}$.
In bulk elemental superconductors, the sharp $R(T)$ reflect the near
simultaneous appearance of a finite amplitude, $|\psi_{0}|$, and
long range coherence of the phase, $\phi$. In low superfluid density
superconductors, however, the amplitude first forms at high temperatures
and phase coherence develops at lower temperatures \cite{Emery&Kivelson-phasechangesTc,KT,BeasleyMooijOrlando,HebardFioryKTtransAl}.
High on a transition quasiparticles transiently form Cooper pairs,
which enhances the quasiparticle or fermionic contribution to the
conductivity, $\sigma_{f}$. These pair or amplitude fluctuations
give rise to the initial drop in $R(T)$ \cite{ALfluctuation}. At
very low values of $R$, a substantial Cooper pair density exists
and the conductivity of these bosons, $\sigma_{b}$, controls $R(T)$.
$\sigma_{b}$ is limited by the motion of vortices which causes fluctuations
in the phase of the cooper pair condensate. In between, a two fluid
model may best describe the transport \cite{a-MOGe-Yazdani&Kapitulnik}. 

Physical interpretations of the $R(T)$ in high sheet resistance films
especially those near the SIT relies heavily on distinguishing the
amplitude and phase fluctuation dominated regimes \cite{Ong-highTcvorticesabove,Maple-highTcvorticesabove,highTc-novorticesabove,InOx-scale}.
Most often, explicit models for $\sigma_{f}(T)$ and $\sigma_{b}(T)$
do not exist as a guide and qualitative arguments must prevail. In
this paper we present the magnetic flux response of the $R(T)$ of
very low superfluid density (high sheet resistance) \cite{criticalampfluct,TAFFbehavior-Jay}
films patterned with a nanoscale array of holes. We use the quality
of the flux response at different points along their transitions to
determine the presence of well-defined vortices \cite{Resnickphaseresponse}
and thus provide insight into where the amplitude fluctuation dominated
transport gives over to phase fluctuation dominated transport on an
$R(T)$. In addition, we propose that this method can be used to directly
detect the existence of vortices in settings where their presence
has been contentious \cite{Goldman-vorticesoninsulatorside}.

\begin{figure}
\includegraphics[%
  width=0.9\columnwidth,
  keepaspectratio]{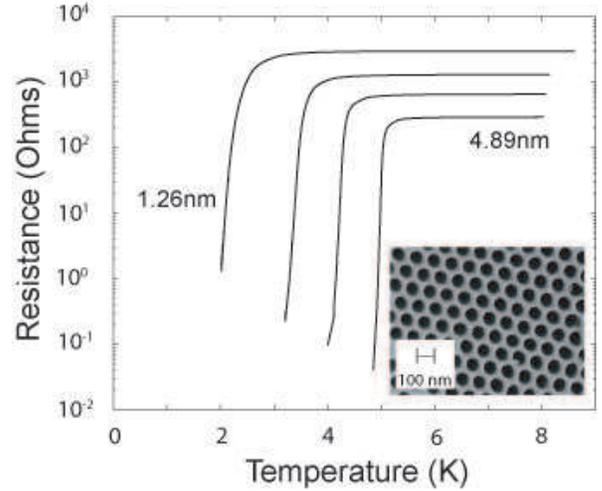}

\caption{Superconducting transitions for a perforated film. Inset shows an
SEM image of one of the substrates used in these experiments. The
hole lattice constant is $a=100$nm and $R_{hole}=33$nm.\label{cap:SIT}}
\end{figure}

As a consequence of phase coherence, magnetic field periodic behavior
is expected in films with multiply connected periodic geometries (see
inset of Fig. \ref{cap:SIT}) \cite{Fiory-scfilmwithholes,oscillationsexpected}.
For example, the resistive transitions of thick Nb films \cite{XuNbpaper}
on nanopore arrays and thick Al films with much larger lattice and
hole dimensions \cite{meanfieldexpression} oscillate in temperature
while maintaining their shape in an applied magnetic field. For holes
separated by segments on the order of $\xi_{0}$ or less \cite{XuNbpaper}\begin{equation}
\frac{\Delta T_{c}}{T_{c0}}=-\left(\frac{\xi_{0}}{a}\right)^{2}\left(\frac{1}{4}-\left(\frac{\phi}{\phi_{0}}-n-\frac{1}{2}\right)^{2}\right)\label{eq:MFT-LP}\end{equation}
 where $\Delta T_{c}$ is the shift in the transition, $\xi_{0}$
is the (dirty) coherence length, $a$ is the hole to hole spacing
\cite{SEMfootnote}, $n$ is an integer, and $\phi$ is the applied
magnetic flux. $\phi_{0}=H_{m}\left(\frac{\sqrt{3}}{2}a^{2}\right)$
is the magnetic flux corresponding to having one flux quantum per
unit cell and defines the unit cell matching field, $H_{M}$. Eq.
\ref{eq:MFT-LP} is derived within the context of mean field theory,
or in the absence of appreciable fluctuations of the superconducting
order parameter and without considering vortex motion. Therefore,
the physical picture which leads to Eq. \ref{eq:MFT-LP} is that of
immobile vortices which penetrate the film through the array of holes.
These vortices generate screening currents which, at $H=nH_{M}$,
nearly cancel in the interior of the film so that $\Delta T_{c}\simeq0$.
At incommensurate fields the screening currents give rise to pair
breaking and thus, a reduction in $T_{c}$. The microscopic picture
described by Eq. \ref{eq:MFT-LP} is the familiar Little-Parks physics.

Eq. \ref{eq:MFT-LP} accounts well for the flux periodic behavior
of thick Nb films \cite{XuNbpaper}. Our data, however, show that
this physical picture is not valid for ultrathin high sheet resistance
films with low superfluid density. Amplitude fluctuations prevent
the formation of screening currents as described above when the fraction
of the normal state resistance $r=R/R_{N}\gtrsim0.5$. In addition,
as $r$ decreases ($r\lesssim0.15$) and screening currents appear,
the films' susceptibility to thermal fluctuations allows activated
vortex motion to dominate the transition. Our data also indicate that
the region of the transition dominated by amplitude fluctuations grows
with decreasing $T_{c0}$ and hence increasing proximity to the SIT.

Our experiments were conducted on homogeneous quench condensed Bi/Sb
films similar to those employed in previous studies of the SIT and
Thermally Activated Flux Flow (TAFF) \cite{criticalampfluct}. These
films have very low superfluid density because they are thin and disordered
\cite{TAFFbehavior-Jay,criticalampfluct}. Two films were made simultaneously
for each experiment. One film was deposited on a fire-polished glass
substrate while the other was deposited on a nano-perforated Anodic
Aluminum Oxide (AAO) substrate (see the inset of Fig. \ref{cap:SIT}).
The latter assumed the honeycomb geometry of the substrate. The preparation
of the AAO substrates can be found elsewhere in the literature \cite{XumakingAAOsubstrates,sciencemakingAAO,makingAAO,makingvariablespacingAAO}.
Those used in these experiments had a hole lattice constant, $a$,
of 100 nm and a hole radius, $R_{hole}=33$ nm. Both substrates were
precoated with 50 nm of thermally evaporated Ge (in an attempt to
smooth the small surface roughness of the AAO substrate) before Au
contact pads were deposited at room temperature. Subsequently both
substrates were mounted in our cryostat where homogeneous Bi films
are fabricated at $T=8$K by first evaporating a thin film of Sb ($<1$nm,
which ensures the films' homogeneous morphology) and then depositing
the desired thicknesses of Bi through sequential depositions. In this
way, a series of Bi films were fabricated without breaking vacuum
or warming.

The $R(T)$ were measured using standard four-terminal, low-frequency
ac techniques and acquired in the regime of applied currents where
the films exhibit an ohmic response. Nonlinearities associated with
the phase transition to the zero resistance state, which are expected
at lower temperatures and have been observed in some wire arrays do
not appear in this experiment's range \cite{Ling-wirearray,Higgins-wirearray,Teitel-jjarray}.
The normal state resistances, $R_{N}$, of neighboring film regions
were compared to assess film homogeneity and found to agree to $<7\%$.
Magnetic fields were applied perpendicular to the planes of the films.
Our thermometry consists of a calibrated carbon glass resistor which
has a negligible magnetoresistance in the range of fields used in
these experiments.

\begin{figure}
\includegraphics[%
  width=0.93\columnwidth]{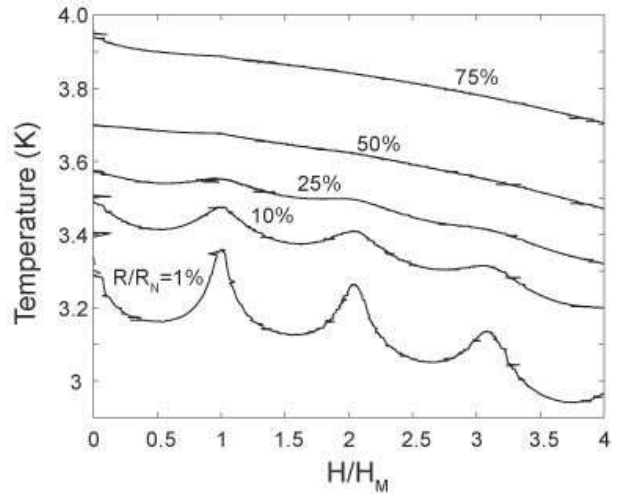}

\caption{Iso-dissipative measurements of temperature vs magnetic field (normalized
by the matching field) for a representative perforated film with $d_{Bi}=1.90nm,$
$T_{co}=3.70$K, $r_{hole}=33$nm, and hole spacing $a=100$nm. Inset
shows the corresponding superconducting transition with the arrows
indicating the value of $R/R_{N}$ where the curves in the main panel
were taken. Eq. \ref{eq:MFT-LP} predicts that each curve should exhibit
the same magnitude oscillation of $\approx0.01$K. \label{cap:HvsT}}
\end{figure}

Data from a series of four superconducting films with $T_{c0}$ ranging
from 2.65 to 5 K and normal state resistances, $R_{N}=R(8K)$, ranging
from 3.3k$\Omega$ to 300$\Omega$ are shown in Fig. \ref{cap:SIT}.
The reference film exhibits the same range of $T_{c0}$ and a similar
increase in the widths of its transitions with decreasing $T_{c0}$.
The latter characteristic has been ascribed to a growing fluctuation
dominated regime \cite{criticalampfluct}. The systematic reduction
in $T_{c0}$ is believed to result from disorder enhanced Coulomb
repulsion effects that grow with increasing $R_{N}$ and possibly
drive the SIT \cite{Belitz-Kirkpatrick,Finkelstein}. 

Despite the breadth of the superconducting transitions they do oscillate
with increasing magnetic field. The amplitude of the oscillations,
however, depends strongly on the reduced resistances, $R(T)/R_{N}$,
at which they are measured. That is, unlike thicker films the shape
of the transitions change in field \cite{XuNbpaper,meanfieldexpression}.
Figure \ref{cap:HvsT} shows a typical series of isodissipative (i.e.
fixed $r$) measurements of $\Delta T$ \cite{isodissipativemst}.
($R_{N}$ does not change in this range of applied fields.) The lower
$r$ curves exhibit oscillations that diminish in size with increasing
field and are superimposed on a nearly quadratic background. The peaks
in the data appear at $H=nH_{m}=n(2145\pm50)$ Gauss where $n$ is
an integer. $H_{M}$ corresponds closely to the calculated value for
one flux quantum per unit cell of the hole array (see \cite{SEMfootnote}).
Thus, these oscillations result from the collective response of the
hole array rather than from the responses of single holes \cite{Abilio&Pannetier}.
A feature of particular interest is that the peak amplitudes are larger
at lower $r$.

\begin{figure}
\includegraphics[%
  width=0.93\columnwidth]{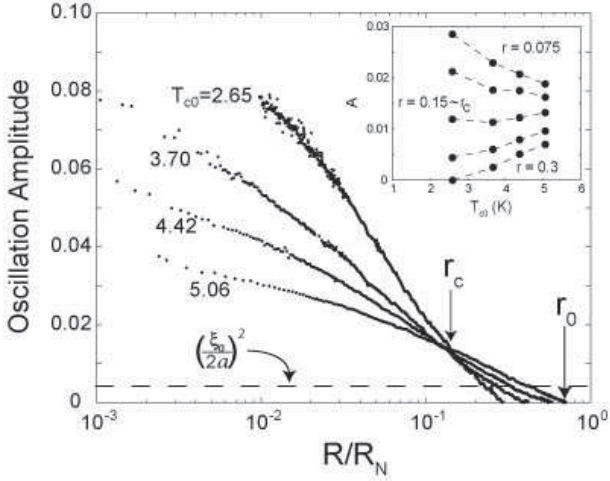}

\caption{Normalized amplitude of the first temperature oscillation as a function
of fractional resistance for 4 perforated films with$2.65<T_{c0}<5.06$K,
$R_{hole}=33$nm, and $a=100$nm. Inset shows the unexpected behavior
of the $A(r)$ with $T_{c0}$ near $r_{c}$.\label{cap:ampvsRandTc}}
\end{figure}

We characterize the flux periodic response using a normalized oscillation
amplitude, A, for the first oscillation:\begin{equation}
A=\frac{T(r,0)+T(r,H_{M})}{2T_{c0}}-\frac{T(r,H_{M}/2)}{T_{c0}},\label{eq:NormAmp}\end{equation}
 Each solid curve of $A$ as a function of $r$ shown in Fig. \ref{cap:ampvsRandTc}
was calculated using individual measurements of $R(T)$ at 0, $H_{m}/2$,
and $H_{m}$. The dashed curve represents the mean field oscillation
amplitude, $A_{MF}=\left(\xi_{0}/2a\right)^{2}$, derived from Eq.
\ref{eq:MFT-LP}. $\xi_{0}\simeq10$ nm represents the average value
for these films as determined from measurements of the upper critical
fields of the reference films. It varied by $~20\%$ over this range
of $T_{c0}$. 

The oscillation amplitudes exhibit a rich dependence on $r$ and $T_{c0}$,
unlike the constant mean field prediction. For all $T_{c0}$, $A$
grows nearly logarithmically from zero with decreasing $r$ with a
dependence that is stronger for the films with lower $T_{c0}$. For
$r\ll1$, the oscillations surpass 10 times the mean field result.
The $r$ at which oscillations first appear, $r_{0}$, is well defined
and is lower for lower $T_{c0}$. Interestingly, the $A(r)$ for different
$T_{c0}$ cross in the vicinity of a single point near $r=r_{c}\simeq0.15$.
The existence of this crossing point is supported by data on another
hole size. Its significance is brought out partially by the inset
in Fig. \ref{cap:ampvsRandTc}, which shows how $A$ depends on $T_{c0}$
at fixed values of $r$. $r_{c}$ delineates the portions of the $R(T)$
in which $A$ increases or decreases with the primary energy scale
characterizing the transition, $T_{c0}$. %
\begin{figure}[b]
\includegraphics[%
  width=0.93\columnwidth]{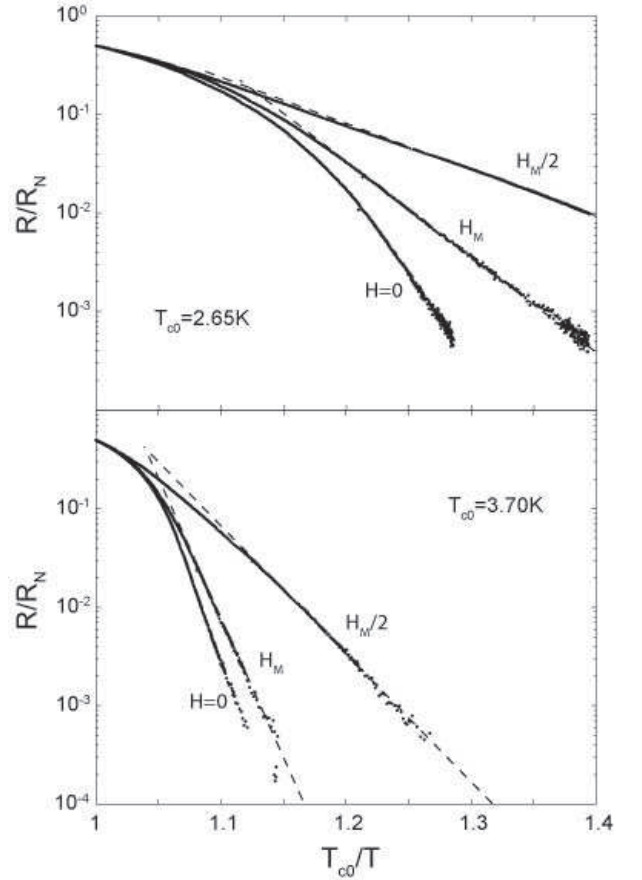}

\caption{Arrhenius plots for the perforated film with $R_{hole}=33nm$. $T_{c0}=2.65K$
(top) and $T_{c0}=3.7K$ (bottom). The slope of the curve characterizes
the energy barrier against vortex motion in the film. The barrier
in the perforated film is non-monotonic with field. Dashed lines are
fits to Eq \ref{eq:TAFF}.\label{cap:ArrheniusPlot}}
\end{figure}

Accounting for the very large $A$ at low $r$ and the vanishing of
$A$ for $r\simeq\frac{1}{2}$ clearly requires consideration of the
effects of fluctuations \cite{Hayashi&Ebisawa}. Unfortunately, detailed
theoretical predictions for the $R(T)$ do not exist. Our identification
of two scales in the data, $r_{0}$ and $r_{c}$, however, provide
a useful guide for the discussion. As we emphasize below, the two
scales define points on the transitions at which the dissipative processes
and character of the fluctuation effects qualitatively change.

Insight into the large amplitude of the oscillations at $r<r_{c}$
comes from investigating the tails of the superconducting transitions
in magnetic field. Figure \ref{cap:ArrheniusPlot} shows Arrhenius
plots of the $R(T)$ collected at $H=0$, $H_{M}/2$, and $H_{M}$
for two films with different $T_{c0}$. The tails of the $R(T)$ follow
\begin{equation}
R(T)=R_{0}e^{(-T_{0}/T)},\label{eq:TAFF}\end{equation}
 which is the signature of Thermally Activated Flux Flow (TAFF). The
energy barrier, $T_{0}$, is lower by more than a factor of 2 at $H=H_{M}/2$
than at $H=H_{M}$. This nonmonotonic variation in $T_{0}$ gives
rise to the $\Delta T$ oscillations that continuously grow as $r\rightarrow0$.
At $H=H_{M}$ the vortices come into registry with the holes and thus
their lattice is made stiff against thermally activated flux flow.
At half integer values of $H/H_{M}$, vortices cannot come into registry
with the holes and simultaneously form a stable lattice. Consequently,
the activation energy is substantially lower at incommensurate fields.
That lower $T_{c0}$ films have larger oscillation amplitudes in this
regime results from the difference in the activation barriers for
different $T_{c0}$ films as seen in Fig. \ref{cap:ArrheniusPlot}.
Even on a normalized temperatuere scale, the lower $T_{c0}$ film
has broader transitions than the higher $T_{c0}$ film. Correspondingly,
the activation energy increases faster than linearly with $T_{c0}$.
This behavior is consistent with the condensation energy dictating
the activation energy scale. A more detailed discussion of the TAFF
behavior will appear in future work.

Moving up the transition to $r>r_{c}$ (Fig. \ref{cap:ampvsRandTc}),
the oscillation amplitudes wane and eventually go to zero. The behavior
as $r\rightarrow r_{0}$ can be attributed to enhanced vortex mobility
that causes them to lose registry with the holes. Nearer $T_{c0}$
the shrinkage of the order parameter makes the barrier to interhole
vortex motion smaller and the likelihood of amplitude fluctuations
greater. Given that the constrictions in the film are quasi-1d (a
few $\xi_{o}$ across) both of these effects will enhance interhole
vortex motion. These effects are more pronounced in lower $T_{c0}$
films since the condensation energy, which is proportional to $T_{c0}$,
governs their strength. Consequently, $A$ is lower in lower $T_{c0}$
films. As $r$ exceeds $r_{0}$, all evidence of well defined vortices
that could be pinned by holes and give rise to a periodic response
disappears. This last regime can be characterized by amplitude fluctuations
that become so strong that they destroy fluxoid quantization altogether.

Thus, even in the absence of a detailed model of $R(T)$ which accounts
for fluctuations in these films, we can identify three distinct fluctuation
regimes defined by $r_{c}$ and $r_{0}$. For $r<r_{c}$, thermally
activated flux motion causes phase fluctuations. In the intermediate
regime, $r_{0}>r>r_{c}$, the vortex mobility is augmented by strong
amplitude fluctuations. And finally, for $r_{0}<r<1$ amplitude fluctuations
dominate and vortices do not exist.

The decrease of $r_{0}$ with $T_{c0}$ implies that the size of the
amplitude fluctuation dominated regime increases closer to the SIT
\cite{criticalampfluct}. This trend can be attributed to the growth
of both the quantum critical regime of the SIT \cite{Kirkpatric-BelitzPRL}
and the classical critical regime \cite{Goldenfeld} as $R_{N}$ increases.
$R_{N}\simeq R_{c}/2$ for the lowest $T_{c0}$ film in the series.
It suggests that the fermionic degrees of freedom strongly influence
the approach to the SIT in uniform films. Experiments on lower $T_{c0}$,
higher $R_{N}$ films will provide insight into whether bosonic (vortices)
or fermionic degrees of freedom dominate the SIT.

We propose that techniques similar to those applied here may be useful
for detecting the existence of vortices in other unconventional situations.
For example, evidence from Nernst Effect \cite{Ong-highTcvorticesabove}
and magnetoresistance \cite{Maple-highTcvorticesabove} measurements
suggests the presence of vortices well above $T_{c}$ in under doped
high $T_{c}$ compounds. Underdoped films patterned with an ordered
array of pinning centers (e.g. holes or magnetic impurities) should
exhibit flux periodic behavior according to our results. Similar arrangements
\cite{Goldman-vorticesoninsulatorside} could test the proposal that
there are vortices on the insulating side of the superconductor to
insulator transition.

In summary, we have quench condensed nano-perforated, homogeneously
disordered Bi films. The regular perforations induce a phase-sensitive
periodic response to an applied magnetic flux and thus provides a
probe capable of distinguishing between the two types of superconducting
fluctuations. In these low superfluid density films, we have found
that the strength of this response crosses over between regions of
the $R(T)$ dominated by fluctuations in the amplitude, $|\psi_{0}|$,
of the superconducting order parameter to that dominated by fluctuations
in the phase, $\phi$. In addition, as $T_{c0}$ is lowered and the
system approaches the quantum critical point of the SIT, the region
of the $R(T)$ dominated by amplitude fluctuations grows.

This work has been supported by the NSF through DMR-0203608, AFRL,
and ONR. We acknowledge helpful conversations with N. Daniilidis.

\bibliography{references}

\begin{thebibliography}{36}
\bibitem[{SEM()}]{SEMfootnote}
\bibinfo{note}{From SEM images we have determined the hole lattice constant
  $a=100\pm5$ nm and thus $H_{M}=2390\pm 240$ Gauss. The position of the peaks
  in the data occur at $H=2145\pm 50$ Gauss in agreement with that calculated
  from the lattice constant. $H_{M}=2145$ Gauss is the normalization used on
  the data throughout this paper.}
\end{thebibliography}

\end{document}